\documentclass[11pt,a4paper]{article} 
\usepackage{jheppub}

\usepackage{amsmath, amssymb}
\usepackage{concmath, palatino}
\usepackage{mathrsfs}

\usepackage{graphicx,epsfig}
\usepackage{epic}
\usepackage{color}

\newcommand{\be}{\begin{equation}}
\newcommand{\ee}{\end{equation}}
\newcommand{\ba}{\begin{eqnarray}}
\newcommand{\ea}{\end{eqnarray}}

\newcommand{\ads}{$AdS_5\times S^5$\ }

\newcommand{\J}{\mathcal{J}}
\newcommand{\SSS}{\mathcal{S}}

\def\Xint#1{\mathchoice
   {\XXint\displaystyle\textstyle{#1}}%
   {\XXint\textstyle\scriptstyle{#1}}%
   {\XXint\scriptstyle\scriptscriptstyle{#1}}%
   {\XXint\scriptscriptstyle\scriptscriptstyle{#1}}%
   \!\int}
\def\XXint#1#2#3{{\setbox0=\hbox{$#1{#2#3}{\int}$}
     \vcenter{\hbox{$#2#3$}}\kern-.5\wd0}}

\def\dashint{\Xint-}

    \newcommand{\beq}{\begin{equation}}
    \newcommand{\eeq}{\end{equation}}
    \newcommand\beqa{\begin{eqnarray}}
    \newcommand\eeqa{\end{eqnarray}}

\newcommand{\CC}{{\mathcal C}} 
\newcommand{\sql}{{ \sqrt\lambda}}
\newcommand{\E}{{\mathcal E}}

\def \la {\label}
\newcommand{\rf}[1]{(\ref{#1})}
\def\ov{\over}
\def\no{\nonumber} \def \aa {{\rm a}}

\def \K  {{ q}}

\subheader{\hfill \rm Imperial-TP-AAT-2012-01}

\title{Leading quantum correction \\ to energy of 
``short'' spiky strings}

\author[a]{Matteo Beccaria } 
\author[a]{, CarloAlberto Ratti } 
\author[b, 1]{, Arkady A. Tseytlin\note{Also at Lebedev Institute, Moscow.}}

\affiliation[a]{Dipartimento di Fisica, Universita' del Salento \& INFN, \\
                     Via Arnesano, 73100 Lecce, Italy} 
                                         
\affiliation[b]{The Blackett Laboratory, Imperial College,  \\
                       London SW7 2AZ, U.K.} 

\emailAdd{matteo.beccaria@le.infn.it}
\emailAdd{carloalberto.ratti@le.infn.it}
\emailAdd{tseytlin@imperial.ac.uk}

\abstract{
We consider semiclassical quantization  of  spiky strings spinning in $AdS_3$ part of  $AdS_5 \times
S^5$ using integrability-based (algebraic curve) method. In the ``short string'' (small spin)
limit the expansion of string energy starts with its flat-space expression. We compute the leading 
quantum string  correction to  ``short'' spiky  string  energy and  find the explicit form
of the corresponding 1-loop coefficient $\aa_{01}$. It turns out to be rational and 
 expressed in terms of the harmonic sums as functions of the number $n$ of spikes. In the special 
 case of $n=2$  when the spiky string reduces to the  single-folded  spinning string the coefficient 
$\aa_{01}$  takes  the value ($-1/4$) found 
 in arXiv:1102.1040.
 We also consider a similar computation for the $m$-folded string 
and more general spiky string with an  extra ``winding'' number, finding similar 
   expressions for $\aa_{01}$.
 These  results may be useful for  a  description of energies of higher excited states in 
 the quantum 
 $AdS_5 \times S^5$ string spectrum, generalizing earlier discussions of 
  the string counterparts of the  Konishi operator.
 }

\arxivnumber{1234.5678}

\begin{document} \maketitle

\bigskip

\section{Introduction and summary }
\label{sec:intro}

The simplest example of the AdS/CFT duality 
\cite{Maldacena:1997re}  states  the equivalence between the spectrum of the planar $\mathcal
 N = 4$ supersymmetric gauge theory and the spectrum of free closed quantum superstring propagating 
 in  \ads. The gauge-theory spectrum can be described in two equivalent ways:
either as a list of possible energies of  states on $\mathbb R\times
 S^{3}$ (as functions of various quantum numbers and `t Hooft coupling $\lambda$) or as a 
 list of anomalous
  dimensions of conformal primary operators on $\mathbb R^{1,3}$ (determined
   by diagonalization of anomalous dimension matrix for single-trace
    gauge-invariant operators). Similarly, the string spectrum is given 
    by the $AdS_{5}$ energies $E$ of string states on a cylinder $\mathbb
     R\times S^{1}$ (in, {\em e.g.}, a light-cone gauge approach) 
     or can be found 
     from the marginality condition for the corresponding string vertex operators on a 
     plane $\mathbb R^{1,1}$ (by diagonalizing  the 2-d anomalous dimension matrix).

The states can be labeled by the conserved charges $C$, i.e. by 
the five spins $(S_{1,2},  J_{1,2,3})$ corresponding to the 
bosonic subgroup $SO(2,4)\times SO(6)$  
of the  $PSU(2,2 | 4)$ symmetry
group 
as well as by higher hidden  charges. 
The AdS/CFT correspondence then implies that  
\be
E_{\rm gauge}(\lambda, C) = E_{\rm string}(\sqrt\lambda, C)\ ,
\ee
where 
$\frac{\sqrt\lambda}{2\pi} =
 \frac{R^{2}}{2\pi\alpha'}$ is  the \ads string tension.

In the strong-coupling ($\lambda\gg 1$) expansion, one expects that massive 
quantum string states should  probe a near-flat region of \ads
and thus should have $E \sim \sqrt[4]\lambda$  \cite{Gubser:1998bc}.
More generally, considerations based on solving the 2d marginality 
condition \cite{Tseytlin:2003ac}    
perturbatively  in ${1 \over  \sqrt\lambda} \ll 1 $ 
  for fixed charges  suggest \cite{Roiban:2009aa}
   that (up to 
 a possible shift of $E$ by a
constant)
\be\label{1}
E^2 = 2N \sql + b_0 +   { b_1 \over \sqrt \lambda} + { b_2 \over (\sqrt \lambda)^2 }  + ...
\ ,
\ee
where $N$ is the flat-space level  number. 
As was argued in \cite{Tirziu:2008fk,Roiban:2009aa, Roiban:2011fe},  
one  can  attempt to find    quantum string energies 
   by starting with the  semiclassical 
strings with fixed  parameters $\CC= { C \over \sql}$
and then take 
 the {\em ``short''} string limit $\CC \to 0$. Indeed, for quantum strings 
 with {\em fixed} charges $C$ the limit  $\sqrt \lambda \gg 1$  implies 
  $\CC = {C \over \sql} 
 \to 0$. Assuming  commutativity of the limits, 
 that  suggests a  possibility 
  to compute the subleading  terms   in the above  expansion 
 by using the semiclassical  string theory methods. 
 The semiclassical string expansion gives\footnote{Here we
 assume 
  that the 1-loop string correction does not contain 
``non-analytic'' terms  which cannot appear in the vertex operator (2d anomalous dimension)  approach
  \cite{Roiban:2009aa}.
This will be   indeed so  in the cases discussed below  treated in the 
algebraic curve approach.} 
 \be
E =\sqrt{\lambda}\,\E_0(\CC)+\E_1(\CC)+\frac
1{\sqrt{\lambda}}\,\E_{2}(\CC)+\ldots\ .  
\ee 
Replacing  $\CC$ by $C \over \sql$
 and
 re-expanding  in  large $\lambda$ for fixed $C$ 
 one should find     that $E$ takes the form  consistent with (\ref{1}):
 \be
 \label{eq:re-expansion}
 E= \sqrt[4]\lambda \Big(k_1 + {k_2 \over \sql}   +  {k_3 \over (\sql)^2} + .... \Big) 
  \ . \ee
  This ``{\em semiclassical}''  approach was successfully applied to the case of the 
 ``short'' string states representing members of the Konishi multiplet
 \cite{Roiban:2009aa,Gromov:2011de, Roiban:2011fe,Beccaria:2011uz,Gromov:2011bz}, 
 matching the results of the weak-coupling TBA approach extrapolated to strong coupling
\cite{Gromov:2009zb, Frolov:2010wt,Gromov:2011de,Frolov:2012zv}.

It is of obvious interest to extend this  approach to 
other quantum string states. 
The semiclassical analog of the 
Konishi representative $\mbox{Tr}(Z \,D^{2}\,Z )$ in the 
  rank-1 $\mathfrak{sl}(2)$     sector  
is the ``short'' folded spinning string  \cite{Roiban:2009aa,Gromov:2011de,Roiban:2011fe}. 
Beyond this twist 2 state, the $\mathfrak{sl}(2)$ sector contains, in particular, 
the operators of the following schematic form 
\be
\label{eq:gaugeops}
\mathcal O = \mbox{Tr}\left(D^{s_{1}}Z\cdots D^{s_{J}}Z\right),
\ee
where $S = \sum_{i=1}^{J}s_{i}$ is the total spin,  $J$ is the R-charge, and $D$ are 
light-cone projected covariant derivatives. Highest weight states of this form
are expected \cite{Belitsky:2003ys,Belitsky:2006en,Belitsky:2008mg,Kazakov:2004nh,Dorey:2008zy,
Dorey:2008vp} to be dual to various spinning 
 string solutions, like the symmetric ``spiky''  string  in $AdS_{3}$
 found  in \cite{Kruczenski:2004wg}
or its generalization to $AdS_{3}\times S^{1}$ discussed in  \cite{Ishizeki:2008tx}.
\footnote{In the Pohlmeyer-reduced  description these configurations 
can  be viewed as multi-soliton 
solutions of a generalized 
sinh-Gordon model \cite{Jevicki:2008mm,Jevicki:2009uz}.}

\def \c {{\rm c}}

In the weak-coupling gauge theory 
  the operators  (\ref{eq:gaugeops}) correspond to  eigenstates of
   the $\mathfrak{sl}(2)$ 
spin chain of length $J$.
The large $S$ limit  is a semiclassical limit \cite{Belitsky:2003ys}
 with $1/S\sim \hbar$. The limit when all $s_i$ in 
  (\ref{eq:gaugeops}) are large was analyzed   in detail in 
\cite{Dorey:2008zy,Dorey:2008vp,Losi:2010hr}. The corresponding 
  Bethe ansatz solutions are associated with sectors characterized by a positive
  integer $n\le J$. For each $n$, their description  involves 
 a genus $n-2$ algebraic curve $\Gamma_{n}$ with 
 $n-2$ moduli $q_{3}, \dots, q_{n}$ which are higher 
conserved charges of the spin chain. They are quantized according to 
suitable Bohr-Sommerfeld conditions
 making the spectrum discrete.
%

On the semiclassical string theory  side,  it is possible to show that the 
spiky strings 
are finite gap solutions also 
associated with a spectral curve that can be put into 
 one-to-one  correspondence with the gauge theory curve 
 \cite{Dorey:2008zy,Dorey:2008vp}.  Besides, it is possible to determine the  cut structure in few  
  special cases.
  The all-order Bethe ansatz   description of the  spiky string solutions 
  was presented in   \cite{Freyhult:2009bx,Kruczenski:2010xs}.
  
  \def \a  {\alpha} 
  
  Below  we will  be interested in two particular 
  solutions. The first one  is the ``{\em symmetric spiky string}'' solution
  spinning in $AdS_3$ and  having also 
an  angular momentum $J$ in $S^{5}$. This 
is a  configuration with $n$ spikes symmetrically distributed
around the $AdS_3$ center. The second solution  is the 
{\em $m$-folded string}, i.e. 
a  closed   string folded on 
itself $m$ times with spin in  $AdS_{3}$ and orbital momentum in  $S^{5}$ 
\cite{Gubser:2002tv,Frolov:2002av}.
In  flat space ($ds^{2} = -dt^{2}+dx^{2}+dy^{2}$) these solutions are given by 
(in conformal gauge)
\def \s {\sigma}
\ba
\mbox{spiky} &:&
\qquad
\left\{
\begin{array}{lll}
t &=& A\,(n-1)\,\tau  \ ,     \ \ \ \ \ \ \ \ \ \ \ \ \ \ \      \s_\pm\equiv  \tau\pm \sigma \\
x &=& {1 \ov 2} A\Big(  \cos[(n-1)\,\s_+]+\,(n-1)\,\cos\s_- \Big), \\
y &=& {1 \ov 2} A \Big( \sin[(n-1)\,\s_+]+\,(n-1)\,\sin\s_- \Big), \\
E &=&  \sqrt{ {4 (n-1)\ov n \a' } \, S} \ , \ \ \ \ \ \ \  S={n (n-1)  \ov 4 \a'} A^2 
\end{array}\right. \la{sp} \\
\mbox{$m$-folded} &:&
\qquad
\left\{
\begin{array}{lll}
t &=&A\,m\,\tau,\\
x &=& A\,\sin(m\,\sigma)\,\cos(m\,\tau)= 
{1 \ov 2} A \Big( \sin( m \s_+)   +   \,\sin ( m\s_-) \Big)     \, , \\
y &=& A\,\sin(m\,\sigma)\,\sin(m\,\tau)
=
{1 \ov 2} A \Big( -\cos( m \s_+)   +   \,\cos ( m\s_-) \Big) 
\, ,  \\
E &=&  \sqrt{ {2 m\ov \a'}\, S} \ , \ \ \ \ \ \ \ \ \  \ \ \ \ S={m \ov 2 \a'} A^2 \ . 
\end{array}\right. \la{fo}
\ea
One can compute the explicit values of the 
charges $q_{j}$ for the corresponding  two $AdS_5 \times S^5$ 
solutions as was shown in \cite{Dorey:2008vp}. 
The result is particularly simple since most of the 
cuts collapse and we are left with a 2-cut solution, asymmetric in the spiky string 
case. This fact 
is a major technical simplification and it 
also shows that these two  solutions can be viewed 
as direct   generalizations of the single-folded spinning 
 string  (which for $S=J=2$ is  dual to
 the Konishi state in the $\mathfrak{sl}(2)$ sector). 
 
 It is then natural to try  to compute the  leading quantum string corrections 
 to  the  corresponding ``{\em short}'' string energies.
 Instead of attempting the direct   semiclassical quantization of the 
 superstring,   one may utilize the integrability 
 by applying  the algebraic curve approach developed, in particular,   in 
 \cite{Gromov:2007aq,Gromov:2007ky,Gromov:2008ec}.
 This would   be a direct 
  generalization  of  the 
 1-folded string  computation  presented  in \cite{Gromov:2011de}.

\bigskip
This is the problem that we address in the main part of this  paper.
Our results may  be summarized as follows. 
The  short string expansion of the energy of the symmetric 
spiky string with $n$ spikes, spin $S$ and angular momentum $J$ has the form 
\ba
\label{eq:spiky-expansion}
&&E_{\rm spiky} = \sqrt{4\big(1-\frac{1}{n}\big)\,\sqrt\lambda\,S}\, \Bigg[
1+\frac{1}{\sqrt\lambda}\Big(
\underbrace{
\frac{2\,n^{2}-5\,n+5}{4\,n\,(n-1)}\,S+\frac{n}{8\,(n-1)}\,\frac{J^{2}}{S}}_{\rm classical}+
\underbrace{\aa_{01}^{\rm spiky}(n)}_{\rm quantum}
\Big) \nonumber \\
&&\ \ \ \ \ \ \ \ \ \ \ \ \ \ \ \ \ \ \ \ \ \ \ \ \ \ \ \ \ \  \ \ \ \ \ \ \ \ \ \ \ \ \ 
\ \ \ \ \ \ \ \ \ \ \ \ \ \ \ \ \ \ \ \ \ \ \ \ \ \ \ \ \ \ +\ \mathcal O\big(\frac{1}{(\sqrt\lambda)^{2}}\big)
\Bigg]\ ,
\ea
where the ``{\em classical}'' terms come from the small spin  expansion of the classical spiky 
string energy \cite{Kruczenski:2004wg}   while
$\aa_{01}^{\rm spiky}(n)$ encodes our result for the leading 
one-loop string correction.
The case of the folded string is $n=2$.
The corresponding expression for the square of the energy has a more transparent structure 
consistent with (\ref{1})
 (as explained above, here  we assume that $S,J,n$ are fixed while $\lambda \gg 1$) 
\be\label{5}
 E^{2}_{\rm spiky} = 4 \big(1-\frac{1}{n}\big)\, \sqrt\lambda \, S\,\Big[
1+ \frac{2}{\sqrt\lambda}
 \,\aa_{01}^{\rm spiky}(n) 
+\mathcal O\big(\frac{1}{(\sqrt\lambda)^{2}}\big) 
\Big] +  J^2 +  4 \Big( 1- {5\over 2n} + {5\over 2n^2}\Big)\, S^2
+ ...  \ . 
\ee
The similar  expansion of the energy of the $m$-folded string  reads
\be\label{6}
E_{\rm folded} = \sqrt{2\,m\,\sqrt\lambda\,S}\,\Bigg[
1+\frac{1}{\sqrt\lambda}\Big(
\underbrace{
\frac{3}{8\,m}\,S+\frac{J^{2}}{4\,m\,S}}_{\rm classical}+\underbrace{\aa_{01}^{\rm folded}(m)}_{\rm quantum}
\Big)+\mathcal O\big(\frac{1}{(\sqrt\lambda)^{2}}\big)
\Bigg],
\ee
leading to 
\be
 E^{2}_{\rm folded} =  2\,m\, \sqrt\lambda \, S\,\Big[
1+\frac{2}{\sqrt\lambda}\,\aa_{01}^{\rm folded}(m)
+\mathcal O\big(\frac{1}{(\sqrt\lambda)^{2}}\big)
\Big]  + J^{2}+   {3\over 2 } S^2  +   ...  \ . \label{55}
\ee
The standard  folded  string case is $m=1$, i.e. 
\be 
E_{\rm spiky}(n=2) = E_{\rm folded}(m=1)   \ . \label{678}
 \ee
Our main results are   the  closed expressions for  the one-loop 
 contributions $\aa_{01}^{\rm spiky}(n)$ and $\aa_{01}^{\rm folded}(m)$~\footnote{Note that the
    leading 
quantum correction $\aa_{01}$ is thus  rational.
 This short string correction should be  captured by the asymptotic
Bethe Ansatz equations (``wrapping'' contributions should  start to appear 
at $\mathcal O(S^{2})$ order, cf. \cite{Basso:2011rs}). Analogous coefficients appearing at higher loop orders
are expected to be rational combinations of $\zeta$-numbers with increasing transcendentality. An example is 
the $\zeta_{3}$ term in the $\mathcal O(\lambda^{-3/4})$
coefficient  for the energy of the Konishi operator~\cite{Gromov:2011bz}. For the spiky and $m$-folded string, these coefficients will be generically dependent on the number of spikes and $m$.
}:
\ba
\aa_{01}^{\rm spiky}(n) &=& -\frac{1}{8}+\frac{1}{2}\,\K(n-1)\ ,\label{s} \\
\aa_{01}^{\rm folded}(m) &=& \,\K(m)\ , \label{f}
\ea
where   the function $\K(r)$  is given by the following combination of the  harmonic sums~\footnote{Note
that  the harmonic sum  can be written in terms of the logarithmic derivative of the Gamma function, 
$\psi(z) = 
\frac{d}{dz}\log\Gamma(z)$, as follows: 
$H_{r} = \psi(r+1)+\gamma_{\rm E}$.
}
\ba
&&\K(r) = -\frac{3}{4r}+   {2} H_{r} -H_{2r}\ ,\label{k} \\
&&H_{r} \equiv  \sum_{\ell=1}^{r}\frac{1}{\ell} \ , \ \ \ \ \ \ \ \ \ H_0 =0 \ . \label{h}
\ea
In the  special case of $n=2$ or $m=1$  corresponding 
to the  1-folded spinning string 
 we find, in agreement with \cite{Gromov:2011de}   
\be 
\aa_{01}^{\rm spiky}(2) = \aa_{01}^{\rm folded}(1)= \K(1)=- {1 \over 4}  \  . \label{ff}
\ee
Note that for positive integers $m,n >1 $ one has 
$ 2m > 4 (1 - {1 \over n})$ so that the spiky string with same spin $S$ as the 
multifolded  string with $m>1$
has lower energy -- it corresponds  to a state on
a lower flat space string level
(the single-folded string has, of course lower energy 
than the spiky string with $n >2$). 

The matching of  ``{\em short}'' string  energies and anomalous dimensions of 
finite-length gauge theory operators requires  a precise 1-1 mapping of the string   states and the 
 operators.
This is non-trivial as remarked in \cite{Dorey:2008zy}.  Indeed, the 
 bound $n\le J$ holds in the perturbative  gauge theory, but  does not appear to be
 present on the semiclassical string theory side.
Recently, the  mirror TBA equations of \cite{Arutyunov:2009ax} were   solved numerically 
starting at weak coupling 
for several two-particle states 
dual to ${\mathcal N} = 4$ SYM operators from the $\mathfrak{sl}(2)$ sector with various values of the charge $J$
 and mode number $n$ \cite{Frolov:2012zv}. The interpolation of the  results
to strong coupling 
 for low values of $n$ does not appear to  agree with the explicit 
 values of our re-expansion  (\ref{eq:re-expansion}). This is not totally surprising since moderate values of $n$ at fixed $J$ fall in the 
 {\em middle} of the allowed range $n\le J$ whereas on  the semiclassical
string theory  side $n$ is unbounded.\footnote{The  value $n=1$ is
 special  as  this  is the minimal value corresponding to the ground state.
 This can be an  explanation of the success of the  matching  in the case of the Konishi operator.}

Nevertheless, it is of interest to try to  apply (\ref{5}) to a
simple 
generalization of the   Konishi operator
($S=J=2$)  in the $\mathfrak{sl}(2)$ sector. 
This is   a particular  highest weight 
 state with $S = J = 3$ that   can be  schematically represented by an operator 
\be
\mathcal O_{3} = \mbox{Tr}(D Z\,DZ\,DZ)+\cdots,
\ee
where dots stand for different distributions 
of the three covariant derivatives such that $\mathcal O_{3}$
is an eigenstate of the dilatation operator.
 The precise identification of the 
string state dual to this operator 
 requires  detailed analysis  
which is beyond the scope of this work. 
Assuming that it corresponds to  the $n=3$ spiky string we then 
find from (\ref{5}) the following prediction   for the 
strong-coupling expansion of its anomalous dimension 
\ba
E_{\rm spiky} (S=J=n=3) &=& 2\,\sqrt{2}\,\sqrt[4]\lambda\,\Big[1+
\frac{41}{24}\,\frac{1}{\sqrt\lambda}
+\mathcal O\big(\frac{1}{(\sqrt\lambda)^{2}}\big)\Big]\ .
\ea
For comparison,  $E_{\rm folded} (S=J=3, m=1) =
 \sqrt{6}\,\sqrt[4]\lambda\,\Big[1+
\frac{13}{8}\,\frac{1}{\sqrt\lambda}
+\mathcal O\big(\frac{1}{(\sqrt\lambda)^{2}}\big)\Big]$
is lower.

Let us   mention  also
the limit of  large number of spikes $n\to\infty$.
In the ``{\em long}'' string case 
 it is possible to scale $E, S, J$ with $n$ keeping
$\frac{E+S}{n^{2}}$,
 $\frac{E-S}{n}$, and $\frac{J}{n}$  fixed \cite{Ishizeki:2008tx}. 
 In this limit the spiky string approaches 
 the boundary of $AdS_5$;  the corresponding solution 
 can be interpreted as
  describing a periodic-spike string moving in $AdS_3-\mbox{pp-wave} \times  S^1$ background
 \cite{Kruczenski:2008bs,Ishizeki:2008tx}. 
  In the  ``{\em short}'' string case  we are interested here 
  we may take $n $ large while keeping  $S, J$
 fixed. We then 
 find  that while 
   the classical contribution in (\ref{eq:spiky-expansion}) has a finite limit,   the one-loop
 correction goes as $\log n$~\footnote{That may be 
 implying  that such a limit is not well-defined in the 
 strong-coupling expansion.}   ,
 \be
 E \stackrel{n\to\infty}{=} 2\,\sqrt{\sqrt\lambda\, S}\,\Bigg[
1+\frac{1}{2\sqrt\lambda}\Big(
\underbrace{ \,S+\frac{J^{2}}{4S}}_{\rm classical}+
\underbrace{
\log\frac{n\,e^{\gamma_{\rm E}}}{2}    -
 \frac{1}{4}+\mathcal O\big(\frac{1}{n}\big)
}_{\rm quantum}
\Big)+\mathcal O\big(\frac{1}{(\sqrt\lambda)^{2}}\big)
\Bigg] \ . 
 \ee

 \bigskip\noindent
As was mentioned above, the symmetric spiky string and
 the $m$-folded string are special in the sense that
the corresponding  algebraic curve reduces to a curve with two cuts
 on the real axis on opposite sides of the origin. 
Such a solution is fully characterized by the positive 
mode numbers $-n_{L}$ and $n_{R}$ associated with the left
 and right cuts. These mode numbers appear in the corresponding  flat-space  solution 
 that generalizes  both  \rf{sp} and \rf{fo}.
   The symmetric spiky string solution has
  $(n_{L}, n_{R}) = (1,n-1)$ while the $m$-folded string has
$(n_{L}, n_{R}) = (m, m)$. 
The  $(n_{L}, n_{R})$ solution in \ads  describes 
 a generalised symmetric 
spiky string with a  possible $AdS_{3}$ winding discussed in \cite{Dorey:2008vp}
(see also Section (5.1.3) in 
\cite{Losi:2010hr}).
The corresponding expression for the 
one-loop corrected energy  is the following generalization of both 
\rf{eq:spiky-expansion},\rf{5}  and \rf{6},\rf{55}
\ba
&& E_{(n_{L}, n_{R})} = \sqrt{\frac{4n_{L}\,n_{R}}{n_{L}+n_{R}}\, \sqrt \lambda\, S}\,
\Big[ 1+   {1 \over \sqrt \lambda} ( \aa_{01} (n_{L}, n_{R})  + ...) + ...  \Big] \ , \label{87}\\ 
&&E^2_{(n_{L}, n_{R})} = 
{\frac{4n_{L}\,n_{R}}{n_{L}+n_{R}}\, \sqrt \lambda\, S}\,
\Big[ 1+   {2 \over \sqrt \lambda} \aa_{01} (n_{L}, n_{R})  
+\mathcal O\big(\frac{1}{(\sqrt\lambda)^{2}}\big) 
\Big] + J^2 + \gamma   S^2 + ...  \label{77} \\
&& \aa_{01} (n_{L}, n_{R}) = {1 \ov 2} \Big[\K(n_{L}) \, + \,\K(n_{R})\Big] 
 \ , \label{777}
\ea
where $\K$ was defined in \rf{k}. 
Indeed, as one can readily see  (cf. \rf{s},\rf{f},\rf{ff}) 
\be
\aa_{01}^{\rm spiky}(n) = \aa_{01} (1, n-1) \ , \ \ \ \ \ \ \ \ \ \ \ \ \ \ \
\aa_{01}^{\rm folded}(m) = \aa_{01} (m, m) \ . \label{9}
\ee
The ``additive'' structure of $\aa_{01}$ in \rf{777} 
 implies that to this low order of the 
``short-string''(or ``near-flat-space'')  expansion the contributions of the ``left'' 
and ``right'' modes simply add up, which should be a consequence of the  integrability.

 \bigskip\noindent
 The structure of the rest of this paper is as follows.
 We shall start in section 2 with a review of the  algebraic curve description of the 
 spiky string. In section 3 we shall  compute the general expression for the 
 1-loop correction to 
 its energy using
 the algebraic curve approach. 
 In section 4 we shall consider the ``short'' string (small spin) expansion 
 of this 1-loop correction  finding the 1-loop coefficient 
 in \rf{s}. In section 5 we shall repeat the same  analysis   for the $m$-folded string
 leading to \rf{f} and present an explanation for the 
 close relation  between the two expressions in \rf{s} and in \rf{f}
 implied by \rf{777}.
 Appendices A and C   contain some technical details and in Appendix B  we present 
 for completeness  the expressions for the  1-loop energy of the spiky and
 $m$-folded strings in the opposite ``long-string'' (large spin) limit.

\section{Algebraic curve  description of the symmetric  spiky string }
\label{sec:2-cut}


As we have mentioned above, starting with the Bethe Ansatz equations, 
 the large-spin  symmetric  spiky string can be described by taking a thermodynamic limit, 
 which leads to a 2-cut solution.  
The  filling fractions associated with the two cuts can be traded for the two constant mode numbers along the 
cuts (see, e.g., \cite{Beisert:2003ea}).
These mode numbers can be identified by analyzing the relevant string solution in flat space
which is a combination of  right and left moving modes. 
For instance, the folded string with $m=1$ has the same number of right and 
left moving excitations, 
i.e. $n_{L} = n_{R}$. 
This suggests that in this case 
the two cuts should  be  symmetric with  modes $\pm 1$ \cite{Beisert:2003ea}. 
Following the same logic, in  the spiky string case 
\cite{Kruczenski:2004wg}  one should get 
 the two asymmetric  cuts with  mode numbers $n_{L} = 1$ and $n_{R} = n-1$, where $n$ 
 is  the number of spikes. This  input data
 was used in the discussion of the large $S$ spiky strings in the context of  the 
all-loop Bethe Ansatz equations \cite{Freyhult:2009bx}. 
The strong coupling limit of the Bethe Ansatz equations  
 \cite{Kruczenski:2010xs} is  the starting point for the construction
of the algebraic curve needed for  the semiclassical expansion \cite{Gromov:2011de}.

According to \cite{Kruczenski:2010xs}, after a suitable rescaling, these equations reduce
to\footnote{Below we use the notation
 $\mathcal{E} = E/\sqrt\lambda$,\ \ $\mathcal{S} = S/\sqrt\lambda$,\ \ $\mathcal{J} = J/\sqrt\lambda$.}
\be\la{ka}
\dashint dx'\,\rho(x')\,\frac{1-\frac{1}{xx'}}{(x-x')(1-\frac{1}{x'^{2}})} = \pi\,n(x)\left(1-\frac{1}{x^{2}}\right)
-\frac{2\pi}{x}\,\mathcal J.
\ee
Here the density $\rho(x)$ describes the momentum carrying roots (there are $S\to\infty$ of them)
and is supported on a certain union of real cuts. The mode number function $n(x)$ is piece-wise constant on the cuts.
It is convenient to define 
\be
\widetilde\rho(x) = \frac{x^{2}}{x^{2}-1}\,\rho(x),\qquad
\widetilde n(x) = n(x)-\frac{2x}{x^{2}-1}\,\mathcal J.
\ee
Then \rf{ka}  becomes 
\be \la{23}
\dashint dx'\,\frac{\widetilde\rho(x')}{x-x'}\left(1-\frac{1}{xx'}\right) = \pi\,\widetilde n(x)\left(1-\frac{1}{x^{2}}\right).
\ee
The zero momentum and normalization conditions are
\be\la{24}
\dashint dx\,\frac{\widetilde \rho(x)}{x} = 0, \qquad
\dashint dx\,\widetilde \rho(x)\,\left(1-\frac{1}{x^{2}}\right) = 4\,\pi\,\mathcal S.
\ee
Finally, the energy can be expressed as
\be\la{25}
\E-\SSS-\J = \frac{1}{2\pi}\dashint dx\, \frac{\widetilde\rho(x)}{x^{2}}.
\ee

\subsection{Solution of the integral  equation in terms of a  resolvent}

Let us focus on the 2-cut case
\be
C_{1} = (d,c),\qquad C_{2} = (b,a), \qquad d<c<-1, \qquad 1<b<a,
\ee
and define the function~\footnote{We choose the standard cut for the square roots.}
\be
\label{eq:f}
f(z) = \sqrt{z-a}\,\sqrt{z-b}\,\sqrt{z-c}\,\sqrt{z-d}\ .
\ee
The resolvent $G(z)$ is defined as  
\be
G(z) = \frac{1}{\pi}\int_{C_{1}\cup C_{2}} dx' \frac{f(z)}{f(x'+i\varepsilon)}\,\frac{\widetilde n(x')}{x'-z}.
\ee
By standard manipulations, one can show that the resolvent obeys
\be
\label{eq:boundary}
G(x\pm i\,0) = \pm\widetilde \rho(x)+i\,\widetilde n(x),\qquad x\in C_{1}\cup C_{2},
\ee
with the explicit expression for $\widetilde\rho(x)$
\be
\widetilde \rho(x) = \frac{1}{\pi}\dashint dx' \,\mbox{sign}(xx') \left|\frac{f(x)}{f(x')}\right|\,
\frac{\widetilde n(x')}{x'-x}.
\ee
The main result is that $\widetilde\rho(x)$ satisfies \rf{23}
 provided 
\be
\label{eq:resolvent-constraint}
G(0) = G_{0} = G_{1} = 0,
\ee
where the constants $G_{0}$, $G_{1}$ are extracted from 
\ba
G(z) &\stackrel{z\to 0}{=}& G(0)+H_{1}\,z+\cdots, \\
G(z) &\stackrel{z\to \infty}{=}& G_{0}\,z+G_{1}+G_{2}\frac{1}{z}+\cdots,
\ea
The zero momentum condition in \rf{24} 
is then automatically satisfied 
and the charges in \rf{24}, \rf{25}  are given by 
\be
\SSS =  -\frac{i}{4}(G_{2}+H_{1}) , \qquad
 \E-\SSS-\J = \frac{i}{2}\,H_{1}.
\ee

\subsection{Special case of the  spiky string}

For the  string with $n$ symmetric spikes, we are to  choose the mode number function as 
\be
n(x) = \left\{\begin{array}{lc}
-1, & \ \  d<x<c<-1, \\
n-1,& \ \   1<b<x<a,
\end{array}\right.
\ee
and then the  resolvent turns out to be 
\ba
G(z) &=& -\frac{2\,i\,n}{\pi\,\sqrt{(a-c)(b-d)}}\,\frac{f(z)}{(b-z)(z-c)}\,\left[
(b-c)\,\Pi\left(\frac{(a-b)(z-c)}{(a-c)(z-b)},r\right)+(z-b)\,\mathbb K(r)
\right]  \nonumber \\
&& -\ i-2\,i\,\mathcal J\,\left[\frac{z}{z^{2}-1}+\frac{f(z)}{2}\left(\frac{1}{f(1)(1-z)}-\frac{1}{f(-1)(1+z)}
\right)\right],\la{gg}
\ea
where $r \equiv  \frac{(a-b)(c-d)}{(a-c)(b-d)}$. The constants appearing 
in the asymptotic expansion of the resolvent 
at $z=0$ and $z=\infty$ are collected in Appendix A.

\section{Semi-classical quantization of the spiky string\\   in the  algebraic curve approach}

A review of the  algebraic curve 
 approach can be found, e.g.,  in  \cite{SchaferNameki:2010jy}
  whose notation we shall follow.

\subsection{Classical data}

The algebraic curve is characterized by a set of quasi momenta $\widehat p_{i}, \widetilde p_{i}$, $i=1, 2, 3, 4$.
They enter the monodromy matrix of the Lax connection for the 
integrable dynamics of the classical \ads superstring 
that has the eigenvalues
\be
\{e^{i\,\widehat p_{1} }, e^{i\,\widehat p_{2} }, e^{i\,\widehat p_{3} }, e^{i\,\widehat p_{4} } |
e^{i\,\widetilde p_{1} }, e^{i\,\widetilde p_{2} }, e^{i\,\widetilde p_{3} }, e^{i\,\widetilde p_{4} }\}.
\ee
These  eigenvalues are roots of the characteristic polynomial and define an 8-sheeted Riemann surface. The classical
algebraic curve has macroscopic cuts connecting various pairs of sheets. Around each cut, we have 
\be
p^{+}_{i}-p^{-}_{j} = 2\,\pi\,n_{ij},\qquad x\in\mathcal C^{ij}_{n},
\ee
where $n_{ij}$ is an integer associated with a given 
 cut. The possible combinations of sheets (a.k.a. polarizations) that are relevant for \ads are
\be
i=\widetilde 1, \widetilde 2, \widehat 1, \widehat 2,\qquad
j=\widetilde 3, \widetilde 4, \widehat 3, \widehat 4.
\ee
The general properties of quasi-momenta are reviewed in \cite{SchaferNameki:2010jy}. In our case, 
similar to \cite{Gromov:2011de}, the non-trivial quasi-momentum is $\widehat p_{2}$ with all other 
being determined by the cut structure, inversion relations, and Virasoro constraints, precisely as in \cite{Gromov:2011de}.
In terms of the resolvent, it can be shown that the required relation is
\be
p_{\widehat 2}(x) = \pi\,\left(-i\,G(x)+2\,\mathcal J\,\frac{x}{x^{2}-1}\right).
\ee

\subsection{Off-shell fluctuations}

To compute the  one-loop correction to string 
energy  in the algebraic curve approach  one is first to find 
 normal mode frequencies for fluctuations around the classical solution. 
 One is then to  quantize the associated
set of effective oscillators and evaluate  the energy correction 
by simply summing up zero-point energies, taking into account the 
Fermi-Bose statistics.

The key  ingredient are the so-called off-shell fluctuation  energies $\Omega^{ij}(x)$, where
$i, j$ label two sheets of the algebraic curve.
These are functions of the spectral parameter $x$ which are actually simpler than the 
normal mode physical frequencies. Off-shell fluctuation energies reduce to them at special non-trivial
points $x^{ij}_{k}$ associated with the $i$-th and $j$-th sheets that  satisfy
\be
p_{i}(x^{ij}_{k})-p_{j}(x^{ij}_{k}) = 2\,\pi\,k.
\ee
Here, $k$ can be regarded as a mode number for the $k$-th normal mode frequency 
with polarization $(i,j)$, i.e. 
$\Omega^{ij}(x^{ij}_{k})$.

Due to the symmetries of the classical solution, the 8+8 bosonic and fermionic frequencies can be written in terms
of only two independent off-shell fluctuations \cite{Gromov:2008ec}. The result in our case is 
\ba
\Omega_{S}(x) &\equiv& {\Omega^{\widetilde 2\,\widetilde 3}(x) = 
\Omega^{\widetilde 2\,\widetilde 4}(x) = 
\Omega^{\widetilde 1\,\widetilde 3}(x) = 
\Omega^{\widetilde 1\,\widetilde 4}(x)  }, \\
\Omega_{A}(x) &\equiv& {\Omega^{\widehat 2\,\widehat 3}(x)}.
\ea
All other frequencies are given by the following expressions 
\ba
\Omega^{(1)}(x) &=& {\Omega^{\widehat 1\,\widehat 4}(x) }= -\Omega_{A}\left(\frac{1}{x}\right)-2, \no \\
\Omega^{(2)}(x) &=& {\Omega^{\widehat 1\,\widehat 3}(x) =\Omega^{\widehat 2\,\widehat 4}(x) }
= \frac{1}{2}\,\Omega_{A}(x)-\frac{1}{2}\,\Omega_{A}\left(\frac{1}{x}\right)-1, \no \\
\Omega^{(3)}(x) &=& {\Omega^{\widehat 2\,\widetilde 3}(x) = \Omega^{\widehat 2\,\widetilde 4}(x) = 
\Omega^{\widehat 3\,\widetilde 1}(x) = \Omega^{\widehat 3\,\widetilde 2}(x)} = \frac{1}{2}\,
\Omega_{A}(x)+\frac{1}{2}\,\Omega_{S}(x), \\
\Omega^{(4)}(x) &=& {\Omega^{\widehat 1\,\widetilde 3}(x) = \Omega^{\widehat 1\,\widetilde 4}(x) = 
\Omega^{\widehat 4\,\widetilde 1}(x) = \Omega^{\widehat 4\,\widetilde 2}(x) }= \frac{1}{2}\,
\Omega_{S}(x)-\frac{1}{2}\,\Omega_{A}\left(\frac{1}{x}\right)-1.\no 
\ea
Thus, we just need to compute the two frequencies $\Omega_{A,S}(x)$ as in the folded string case. A simple 
calculation gives 
\ba
\Omega_{A}(x) &=& -\frac{2}{1+f(0)}\,\left(1-\frac{f(x)}{x^{2}-1}\right)\ , \\
\Omega_{S}(x) &=& \frac{1}{1+f(0)}\,\frac{f(1)\,(x+1)-f(-1)\,(x-1)}{x^{2}-1},
\ea
where $f(x)$ has been defined in (\ref{eq:f}).

\subsection{One-loop correction to the energy}

The one-loop  energy is given, in  general, by 
\be
\label{eq:one-loop-correction}
\delta E^{\rm 1-loop} = \frac{1}{2}\,\sum_{ij}(-1)^{F_{ij}}\oint\frac{dx}{2\,\pi\,i}\,\Omega^{ij}G_{ij},
\ee
where
\be
G_{ij} = \partial_{x}\log\sin\frac{p_{i}-p_{j}}{2}.
\ee
In our case
\ba
\sum_{ij}(-1)^{F_{ij}}\Omega^{ij}G_{ij} &=& 
4\,\Omega_{S}\,G_{\widetilde 2\,\widetilde 3} 
+ \Omega_{A}\,G_{\widehat 2\,\widehat 3} 
+\Omega^{(1)}\,G_{\widehat 1\,\widehat 4} 
+ 2\,\Omega^{(2)}\,G_{\widehat 1\,\widehat 3} \no \\
&&
- 4\,\Omega^{(3)}\,G_{\widehat 2\,\widetilde 4}
 - 4 \,\Omega^{(4)}\,G_{\widetilde 2\,\widehat 4}.
 \ea
As in the case of the folded string \cite{Gromov:2011de}, 
the one-loop correction can be written as 
\be\la{318}
\delta E^{\rm 1-loop} = \delta E^{(1)}+\delta E^{(2)}+\delta E^{(3)},
\ee
where
\ba
\label{eq:three-one-loop}
\delta E^{(1)} &=& \int_{-1}^{1}\frac{dz}{\pi}\,\mbox{Im}(p_{\widehat 2}-p_{\widetilde 2})
\,\partial_{z}\,\mbox{Im}(\Omega_{S}-\Omega_{A}),  \nonumber \\
\delta E^{(2)} &=& \int_{-1}^{1}\frac{dz}{\pi}\,\mbox{Im}\left[
\partial_{z}\Omega_{S}\log\frac{(1-e^{-i\,p_{\widetilde 2}+i\,\overline p_{\widehat 2}})
(1-e^{-i\,p_{\widetilde 2}-i\,p_{\widehat 2}})}{(1-e^{-2\,i\,p_{\widetilde 2}})^{2}} \right.\nonumber \\
&& \left.
-\partial_{z}\Omega_{A}\log\frac{(1-e^{-2\,i\,p_{\widehat 2}})
(1-e^{-i\,p_{\widehat 2}+i\,\overline p_{\widehat 2}})}{(1-e^{-\,i\,p_{\widetilde 2}-\,i\,p_{\widehat 2}})^{2}}
\right], \\
\delta E^{(3)} &=& \frac{2}{1+f(0)}\,\int_{(d,c)\cup(b,a)}\frac{dx}{2\,\pi\,i}\,\frac{f(x)}{x^{2}-1}
\,\partial_{x}\log\sin p_{\widehat 2}.\nonumber
\ea
In the first two integrals we have defined
\be
x = x(z) = z+\sqrt{z^{2}-1}, \qquad \overline p_{\widehat 2} = p_{\widehat 2}\left(\frac{1}{x(z)}\right).
\ee
The third integral is done above the cuts $(d,c)\cup (b,a)$. 

\section{One-loop correction to the energy of the 
short spiky string}

The  above general expressions  allow one to find the  one-loop correction to the spiky string energy.
In particular, in the ``{\em long string}'' (large spin) 
limit one recovers the results found in 
\cite{Freyhult:2009bx}, as we summarize in Appendix \ref{app:long}.
Here we shall concentrate on the 
opposite ``{\em short string}'' limit.


\subsection{Expansion of the classical data}

We define the short string limit as
\be
\mathcal S\equiv  \frac{1}{2}\,s^{2}\to 0,\  \qquad\mbox{with}\qquad \mathcal J = \rho\,\mathcal S \to 0
\ , 
\ \ \ \ \ \ \ \rho\equiv {J \ov S} ={\rm fixed} \ . 
\ee
Then the endpoints of the two cuts have the following expansion 
\ba
a &=& 1+\frac{2 \sqrt{2}}{\sqrt{n\,(n-1)}}\,s+\frac{4}{n\,(n-1)}\,s^{2}+
\frac{n \left(\rho ^2\,n+4\right)+20}{8 \sqrt{2} 
(n\,(n-1))^{3/2}}\,s^{3}+\mathcal O\left(s^4\right), \\
b&=& 1+\frac{1}{8}{\sqrt{\frac{n}{2\,(n-1)^3}} \rho ^2\,
   s^3}-\frac{n \left(n \rho ^2-
   12\right)+20}{256\,\sqrt{2\,n} \,(n-1)^{5/2} 
   }\,\rho^{2}\,s^{5}+\mathcal O\left(s^6\right), \\
c &=& -1-\frac{n\, \rho ^2}{8\, \sqrt{2\,n\,(n-1)}}\,s^{3}+\frac{n
    \left(n \left(\rho ^2+8\right)-28\right)+20}{256 
    \sqrt{2\,n} \,(n-1)^{3/2} }\,\rho^{2}\,s^{5}+\mathcal O\left(s^6\right), \\
d &=& -1-2 \sqrt{\frac{2(n-1)}{n}}
   \, s+\left(\frac{4}{n}-4\right) s^2-\frac{\left(n \left(n \left(\rho 
   ^2+24\right)-44\right)+20\right) s^3}
   {8\,n \,\sqrt{2\,n\,(n-1)}}+\mathcal O\left(s^4\right).
\ea
The classical energy is then given by 
\be
E_{0} = \sqrt\lambda\,\left[
\sqrt{\frac{2(n-1)}{n}} \, s+\frac{\big(n \left[n
   \left(\rho ^2+4\right)-10\right]+10\big) s^3}{8  n \sqrt{2(n-1) n}}+\mathcal O\left(s^4\right)
\right].
\ee
which can be recognized as the classical part in (\ref{eq:spiky-expansion}).

\subsection{The one-loop coefficient $\aa_{01}(n)$}

The 
small $s$ expansion of the one-loop correction starts with 
\be
\label{eq:spiky-normalization}
E_{\rm 1-loop} = \sqrt\frac{2\,(n-1)}{n}\,\aa_{01}^{\rm spiky}\,s + \cdots.
\ee
The  calculation of $\aa_{01}(n)$ 
can be done along the lines explained in \cite{Gromov:2011de}.
The trick 
is to evaluate the one-loop integrals in \rf{318} 
splitting the integration region into  three parts
($\Lambda_{1, 2}$ are auxiliary parameters)
\be
z\in(0,1) = (0,1-s\,\Lambda_{1})\cup(1-s\,\Lambda_{1},1-s^{3}\,\Lambda_{2})\cup
(1-s^{3}\,\Lambda_{2}, 1).
\ee
The integrals can be computed in each interval by a rather straightforward expansion in $s\to 0$. Finally, the three
divergent results are merged and the cutoffs $\Lambda_{1, 2}$ 
can be sent to infinity. This procedure is known as
{\em matched asymptotic expansion} \cite{Methods} (see Appendix \ref{app:asymptotic} for a simple
example). The computation  is rather complicated 
  and finally leads to the following 
results
$$
\begin{array}{c|ccccccccc}
n & 2 & 3 & 4 & 5 & 6 & 7 & 8 & 9 &\cdots \\
\aa_{01}^{\rm spiky}(n) & -\frac{1}{4} &  \frac{7}{48} &  \frac{43}{120} &  \frac{1699}{3360} &  
\frac{3119}{5040} &  \frac{9853}{13860} &  \frac{568291}{720720} &  \frac{2466643}{2882880} &\cdots
\end{array}
$$	
The  closed  expression  for this sequence is 
\be
\label{eq:spiky-closed}
\aa_{01}^{\rm spiky}(n) = -\frac{1}{8}+\frac{1}{2}\,\K(n-1)\ ,
\ee
where 
\be
\K(r) = -\frac{3}{4r}+ {2} H_{r}-H_{2r}\ ,\ \ \ \ \ \ \ \ \ 
H_{r} = \sum_{\ell=1}^{r}{1\over \ell} \ . 
\ee
The analytic expression \rf{eq:spiky-closed} was carefully checked by a 
numerical evaluation of the one-loop
correction extrapolated to $s\to 0$:  
 we verified  the values of $\aa_{01}^{\rm spiky}$
and its independence of $\rho=J/S$   with 
$10^{-8}$ accuracy.

\section{$m$-folded string and relation to the spiky string case}

Let us now consider the folded string case. 
We will  not repeat all the details  since they are completely 
similar to the spiky string case.
The spinning string  solution 
with $m$ folds  
is described by a 2-cut solution with mode numbers
 $\pm m$ \cite{Dorey:2008vp}. The classical energy reads 
\ba
&&E_{0} = m\,\sqrt\lambda\, s \Big(1+\frac{3+2\,\rho^{2}}{16}\,s^{2}+\dots\Big), \\
&& s= \sqrt{2S\ov m \sqrt \lambda  } \ , \ \ \ \ \ \ \ \ \ \ \ \ \rho=\frac{J}{S} \ . 
\ea
The one-loop correction turns out to be  linear in $s\to 0$ 
\be
E_{\rm 1-loop} = m\,\aa_{01}^{\rm folded}(m)\,s+\dots,
\ee
leading to the expression  in Eq.~(\ref{6}). Again,
 the values of $\aa_{01}^{\rm folded}(m)$ can be
determined analytically~\footnote{The first entries of the list have been independently confirmed 
in the recent paper \cite{Beccaria:2012tu}.}
\be
\label{eq:tablesl2}
\begin{array}{c|ccccccccccc}
m & 1 & 2 & 3 & 4 & 5 & 6 & 7 & 8 & 9 & 10 & \cdots \\
\aa_{01}^{\rm folded}(m)  & -\frac{1}{4}& \frac{13}{24} &  \frac{29}{30}&  \frac{2119}{1680} & 
\frac{3749}{2520} &  \frac{23171}{13860} & \frac{658381}{360360} & \frac{2827003}{1441440} 
& \frac{25478473}{12252240} & \frac{508697569}{232792560} & \cdots
\end{array}
\ee
and one can find  a closed formula for
 this sequence\footnote{This requires  analyzing many  more values 
 than  shown in the above  table.}, namely, 
\ba
\aa^{\rm folded}_{01}(m) &=& 
- \frac{3}{4\,m}+ {2} H_{m }-H_{2m} = \K(m) \ . 
\ea
This is the  same function $\K$ as the one that 
appeared  in the spiky string case (\ref{eq:spiky-closed}). 
A possible  explanation of 
the close relation between the  two 1-loop
coefficients 
is a  conjectured decoupling of the left- and right- moving string modes 
 in the short-string  limit.

Indeed, from the algebraic curve point of view the $m$-folded string 
has two symmetric cuts, one with mode number $-m$ and the other with mode 
number $+m$.
On the other hand, the spiky string with $n$ spikes has two asymmetric 
cuts, one with mode number $-1$
and the other with mode number $n-1$. 
We may  assume that,  in the short string limit, the contributions
 from the left and right cuts simply add. 
However, before 
combining them, we have to take into account the different 
normalizations of the cut endpoints. Indeed, 
the short string expansion of the largest  endpoint $a$  of the 
positive cut  reads
\ba
a^{\rm folded} &=& 1+2s+2s^{2}+\cdots, \\
a^{\rm spiky} &=& 1+2\widetilde s+2\widetilde s^{2}+\cdots, \qquad \widetilde s = \frac{\sqrt 2}{\sqrt{n\,(n-1)}}\,s.
\ea
Taking into account the ratio $\widetilde s/s$ and the normalization of $\aa_{01}^{\rm spiky}$ in 
(\ref{eq:spiky-normalization}) this means that in the 
spiky string case  the sum of the contribution of the 
cut with the  mode number 
 $-1$
(half of folded string with  $m=1$) and
of  the cut with the mode number $n-1$ (half of a folded string with 
 $m=n-1$) reads
\be
\aa_{01}^{\rm spiky}(n) =
 \frac{1}{2}\times \, \aa_{01}^{\rm folded}(m=1)+
 \frac{1}{2}\times\,\aa_{01}^{\rm folded}(m=n-1) \ .
\ee
This  is  exactly the relation
(\ref{eq:spiky-closed}) found  in  the spiky string  case.


\section*{Acknowledgments}

We thank N. Gromov for a collaboration during the initial stages of this work and for  very useful 
 discussions of the  results.
We thank D. Serban and D. Volin for 
helping us with  technical details 
of   very involved 
short string analytic computations and also   thank G. Macorini for
 discussions about the  numerical checks of 
the calculation.
We are also grateful to  S. Giombi, M. Kruczenski,  R. Roiban and A. Tirziu for useful 
discussions of related issues of 
semiclassical 1-loop  string  computations. 
The  work of AAT was supported by the  ERC Advanced grant  no. 290456.


\appendix

\section{Expansion of the spiky string 
 resolvent at $z\to 0$ and $z\to \infty$}
\label{app:constants}

The expansion of $G(z)$ in \rf{gg} 
at $z\to 0$ and $z\to \infty$ determines the constants 
 $G(0)$, $G_{0}$, $G_{1}$, $G_{2}$, $H_{1}$ . They were  computed in
  \cite{Kruczenski:2010xs}: 
\ba
G(0) = -i+f(0)\,\left[\frac{2\,i\,n}{\pi}\,\frac{b-c}{bc\sqrt{(a-c)(b-d)}}\,
\overline\Pi+\frac{i\,\mathcal J}{c}
\left(\frac{1-c}{f(1)}+\frac{1+c}{f(-1)}\right)
\right]\ , 
\ea
\ba
G_{0} = \frac{2\,i\,n}{\pi\,\sqrt{(a-c)(b-d)}}\,\mathbb K + i\,\mathcal J\,
\left(\frac{1}{f(1)}+\frac{1}{f(-1)}\right)\ , 
\ea
\ba
G_{1} &=& -i+\frac{2\,i\,n}{\pi}\,\frac{b-c}{\sqrt{(a-c)(b-d)}}\,\Pi + \frac{2\,i\,n}{\pi}
\,\frac{c}{\sqrt{(a-c)(b-d)}}\,\mathbb K\\
&&+\ 
i\,\mathcal J\,
\left(\frac{1}{f(1)}-\frac{1}{f(-1)}\right) \ , \nonumber
\ea
\ba
G_{2} &=& -2\,i\,\mathcal J+\frac{2\,i\,n}{\pi}\frac{b^{2}-c^{2}}{\sqrt{(a-c)(b-d)}}\,\Pi
+\frac{2\,i\,n}{\pi}\frac{(b-c)^{2}(a-b)}{(a-c)\sqrt{(a-c)(b-d)}}\,\Pi'\\
&& +\frac{2\,i\,n}{\pi}
\frac{c^{2}}{\sqrt{(a-c)(b-d)}}\,\mathbb K(r)+i\,\mathcal J\,\left(
\frac{1}{f(1)}+\frac{1}{f(-1)}
\right)-\frac{i}{2}(a+b+c+d)\ , \nonumber
\ea
\ba
H_{1} &=& 2\,i\,\mathcal J+\frac{i}{2}\left(\frac{1}{b}+\frac{1}{c}-\frac{1}{a}-\frac{1}{d}\right)
-\frac{2\,i\,n}{\pi}\frac{(b-c)^{2}(a-b)}{b^{3}c(a-c)\sqrt{(a-c)(b-d)}}\,\overline\Pi'\,f(0)+\nonumber\\
&& -\frac{i\,\mathcal J}{b c}f(0)\,\left(
\frac{(1-c)(1-b)}{f(1)}+\frac{(1+c)(1+b)}{f(-1)}
\right) \ . 
\ea
The coefficients here are defined in terms of  the elliptic functions as follows
\ba
\mathbb K &=& \mathbb K(r), \qquad \Pi = \Pi\left(v, r\right), \qquad
 \overline\Pi = \Pi\left( \frac{c}{b}\, v\, , \ r \right), \\
&&  v\equiv \frac{a-b}{a-c}\ ,
 \qquad\ \ \ \ 
 r \equiv   \frac{(a-b)(c-d)}{(a-c)(b-d)}\ , 
\\
\Pi'(v,r) &=& \frac{\partial\Pi(v,r)}{\partial v} = \frac{1}{2(r-v)(v-1)}\Big[\mathbb
 E(r)+\frac{r-v}{v}\,\mathbb K(r)
+\frac{v^{2}-r}{v}\,\Pi(v,r)\Big]
\ea

\section{Large spin limit of the  spiky string and $m$-folded string}
\label{app:long}

The large spin limit of the spiky and 
$m$-folded string can be computed in a very simple way from our 
expressions for the algebraic curve and adapting the calculation performed in \cite{Gromov:2011de}
for the standard folded string with $m=1$. 
In the notation of that paper, the one-loop correction is 
computed in terms of the three contributions $\delta E^{(1)}$,
 $\delta E^{(2)}$, and $\delta E^{(3)}$
which are the symmetric cases of Eqs.~(\ref{eq:three-one-loop}). Also, we can further split $\delta E^{(3)}$
into an ``{\em anomaly}'' contribution plus a remainder term, i.e.  
$\delta E^{(3)} = \delta E^{(3)}_{an} + \delta E^{(3)}_{m}$. 
The leading and next-to-leading contributions come only from $\delta E^{(1)}$ and $\delta E^{(3)}_{m}$. The other 
terms are suppressed as $\mathcal O(1/\log S)$.

The long string limit $\mathcal S\to \infty$ is achieved when the cut $(b,a)$ endpoints 
have the asymptotic behaviour  $a\to \infty$ and $b\to 1$ with 
\be
\label{eq:endpoints1}
 \SSS= {1 \ov 2\,\pi} \, a\ , \ \ \ \ \qquad \J =
 \frac{1}{\pi}\,\sqrt{b^{2}-1}\,\log\frac{a}{b}\ .
\ee
The one-loop correction to the energy of the ``long''  1-folded string 
turns out to be
\be
\label{eq:long-standard}
E_{\rm 1-loop} = -\frac{3\log 2}{\pi}\log\overline\SSS+\frac{6\log 2}{\pi}+1 + 
\mathcal O\left(\frac{1}{\log \overline\SSS}\right)\ , \ \ \ \ \ \ \ \ \ \ \ \ \ 
\SSS \equiv 8\,\pi\,\SSS \ . 
\ee
The derivation of this result involves 
 fixing the ratio $\ell = J/\log S$ and sending $\ell\to 0$ in the end
 (the 
only important feature is  the 
relation between $J$ and $\log S$ coming from (\ref{eq:endpoints1})).

In the spiky string case, the large spin limit is obtained 
by considering the two cuts $(d,c)\cup(b,a)$ in the limit
\be
d = -u\,a, \ \ \ \ \ \ \   \qquad a\to +\infty,
\ee
where $u$ is a real positive constant. The conditions (\ref{eq:resolvent-constraint}) on the resolvent $G_{0}=G_{1}=0$ give important information
in this limit. We start with $G_{0}$
that admits the large $a$ expansion
\be
G_{0} = \frac{G_{0,-1}}{a}+\mathcal O\left(\frac{1}{a^{2}}\right).
\ee
The vanishing of $G_{0,-1}$ gives the basic relation between $\mathcal J$ and $\log a$
\be
{
\mathcal J\,\left(\frac{1}{\sqrt{(b-1)(1-c)}}+\frac{1}{\sqrt{-(b+1)(c+1)}}\right) = \frac{n}{\pi}\,
\log\frac{16\,a\,u}{(b-c)\,(u+1)}.
}
\ee
Expanding $G_{1}$ we find instead two non trivial terms
\be
G_{1} = G_{1,1}\,a+G_{1,0}+\mathcal O\left(\frac{1}{a}\right).
\ee
The vanishing of $G_{1,1}$ gives
\be
u = \cot^{2}\frac{\pi}{2\,n}.
\ee
The vanishing of  $G_{1,0}$ gives 
\be
\frac{b+c+2}{\sqrt{-(b+1)(c+1)}}+\frac{b+c-2}{\sqrt{(b-1)(1-c)}} = 0
 \quad\longrightarrow\quad b+c=0,
\ee
and the two cuts become symmetric in this limit. 
Using these results to simplify the expression 
for  the spin, we easily obtain the following  leading-order expression
\be
\mathcal S = \frac{n\,a\,\sqrt u}{4\,\pi}+\mathcal O(1).
\ee
If we trade $a$ for  $\mathcal S$ and use $b+c=0$, we can write the relation 
between $\mathcal J$ and $\log \mathcal S$
in the following form (which is a required modification of (\ref{eq:endpoints1}))
\be
\label{eq:modifiedJS}
\J = \frac{n}{2\,\pi}\,\sqrt{b^{2}-1}\,\log\left(\frac{2}{n}\,\sin\frac{\pi}{n}
\,\frac{\overline \SSS}{b}\right).
\ee
From this equation, it is a straightforward exercise to obtain the 
following $n >2$ generalization  of (\ref{eq:long-standard})
\be
\label{eq:long-spiky}
E_{\rm 1-loop}^{\rm spiky} = \frac{n}{2}\,\left[-\frac{3\log 2}{\pi}\log\left(
\frac{2}{n}\,\sin\frac{\pi}{n}\,\overline\SSS\right)+\frac{6\log 2}{\pi}+1\right]+
\mathcal O\left(\frac{1}{\log \overline\SSS}\right),
\ee
which is 
in agreement with the Bethe Ansatz calculation of \cite{Freyhult:2009bx}.
This agreement is,  of course, 
 expected due to the general results of \cite{Gromov:2007ky}. 
 A completely similar treatment
for the $m$-folded string case gives another  $m>1$  generalization 
of (\ref{eq:long-standard})
\be
\label{eq:long-folded}
E_{\rm 1-loop}^{\rm folded} = m\,\left(-\frac{3\log 2}{\pi}\log\frac{\overline\SSS}{m}+\frac{6\log 2}{\pi}+1\right) + 
\mathcal O\left(\frac{1}{\log \overline\SSS}\right).
\ee

\section{On the asymptotic evaluation of 2-scale integrals}
\label{app:asymptotic}

Let us consider the integral 
\be
I(s)=\int_{0}^{1}dx\  f(x) \ , \ \ \ \ \ \ \ \ \ \ \ \ \ \ \ \ \ \ \
f(x) = \frac{1}{(x-1-s)(x-1-s^{3})}\ .
\ee
It can be computed exactly and expanded for $s\to 0$
\be
\label{eq:asympt}
I(s) = \frac{1}{s(s^{2}-1)}\log \frac{s^{2}}{s^{2}-s+1} = 
-2\frac{\log s}{s}-1+\left(\frac{1}{2}-2\,\log s\right) s-\frac{s^{2}}{3}+\mathcal O(s^{3}).
\ee
Let us now describe 
 the general strategy of how to obtain this asymptotic 
 expansion without requiring the knowledge of the
exact integral.  We first split the integral into three parts as
\be
I(s) = (\int_{0}^{1-s\Lambda_{1}}+\int_{1-s\Lambda_{1}}^{1-s^{3}\Lambda_{2}}+
\int_{1-s^{3}\Lambda_{2}}^{1}) \ dx \ f(x) 
\ee
This can be written as 
\be
I(s) =\int_{0}^{1-s\Lambda_{1}}  dx \ f(x)
-s\,\int_{\Lambda_{1}}^{s^{2}\Lambda_{2}}d\tau\ f(1-s\tau)\, 
-s^{3}\,\int_{\Lambda_{2}}^{0}d\xi \ f(1-s^{3}\xi)\, .
\ee
Then we directly 
expand each integrand in powers of $s$ 
\ba
f(x) &=& \frac{1}{(x-1)^{2}}+\frac{s}{(x-1)^{3}}+\frac{s^{2}}{(x-1)^{4}}+\mathcal O(s^{3}), \\
-s\,f(1-s\tau) &=& -\frac{1}{\tau(\tau+1) s}+\frac{s}{\tau^{2}(\tau+1)}+\mathcal O(s^{3}), \\
-s^{3}\,f(1-s^{3}\xi) &=& -\frac{1}{(\xi+1) s}+\frac{\xi s}{\xi+1}+\mathcal O(s^{3}).
\ea
Performing the integrals term by term we find 
\ba
\int_{0}^{1-s\Lambda_{1}}\ dx \ f(x)
 &=& \frac{2-3\Lambda_{1}+6\Lambda_{1}^{2}}{6\Lambda_{1}^{3}}
\,\frac{1}{s}-1+\frac{s}{2}-\frac{s^{2}}{3}+\mathcal O(s^{3}), \\
-s\,\int_{\Lambda_{1}}^{s^{2}\Lambda_{2}}\ d\tau\ f(1-s\tau)
 &=& 
\left(-2\log s+\log\frac{\Lambda_{1}}{\Lambda_{1}+1}-\log\Lambda_{2}-\frac{1}{\Lambda_{2}}\right)\,\frac{1}{s}
 \\
&&+ \left(-2\log s+\log\frac{\Lambda_{1}}{\Lambda_{1}
+1}-\log\Lambda_{2}+\frac{1}{\Lambda_{1}}+\Lambda_{2}\right)\,s
+\mathcal O(s^{3}), \nonumber \\
-s^{3}\,\int_{\Lambda_{2}}^{0}\ d\xi \ f(1-s^{3}\xi) &=& 
\log(\Lambda_{2}+1)\,\frac{1}{s}+\left(\log(\Lambda_{2}+1)-\Lambda_{2}\right)\,s
+\mathcal O(s^{3}).
\ea
Summing up  and sending $\Lambda_{1,2}\to \infty$,
 all the divergences cancel and we recover 
precisely the expansion in (\ref{eq:asympt}).

\bibliography{AC-Biblio}{}

\providecommand{\href}[2]{#2}\begingroup\raggedright\begin{thebibliography}{10}

\bibitem{Maldacena:1997re}
J.~M. Maldacena, {\it {The Large N limit of superconformal field theories and
  supergravity}},  {\em Adv.Theor.Math.Phys.} {\bf 2} (1998) 231--252,
  [\href{http://xxx.lanl.gov/abs/hep-th/9711200}{{\tt hep-th/9711200}}].

\bibitem{Gubser:1998bc}
S.~Gubser, I.~R. Klebanov, and A.~M. Polyakov, {\it {Gauge theory correlators
  from noncritical string theory}},  {\em Phys.Lett.} {\bf B428} (1998)
  105--114, [\href{http://xxx.lanl.gov/abs/hep-th/9802109}{{\tt
  hep-th/9802109}}].

\bibitem{Tseytlin:2003ac}
A.~A. Tseytlin, {\it {On semiclassical approximation and spinning string vertex
  operators in $AdS_{5}\times S^{5}$}},  {\em Nucl.Phys.} {\bf B664} (2003)
  247--275, [\href{http://xxx.lanl.gov/abs/hep-th/0304139}{{\tt
  hep-th/0304139}}].

\bibitem{Roiban:2009aa}
R.~Roiban and A.~A. Tseytlin, {\it {Quantum strings in $AdS_{5}\times S^{5}$:
  Strong-coupling corrections to dimension of Konishi operator}},  {\em JHEP}
  {\bf 0911} (2009) 013, [\href{http://xxx.lanl.gov/abs/0906.4294}{{\tt
  arXiv:0906.4294}}].

\bibitem{Tirziu:2008fk}
A.~Tirziu and A.~A. Tseytlin, {\it {Quantum corrections to energy of short
  spinning string in AdS(5)}},  {\em Phys.Rev.} {\bf D78} (2008) 066002,
  [\href{http://xxx.lanl.gov/abs/0806.4758}{{\tt arXiv:0806.4758}}].

\bibitem{Roiban:2011fe}
R.~Roiban and A.~Tseytlin, {\it {Semiclassical string computation of
  strong-coupling corrections to dimensions of operators in Konishi
  multiplet}},  {\em Nucl.Phys.} {\bf B848} (2011) 251--267,
  [\href{http://xxx.lanl.gov/abs/1102.1209}{{\tt arXiv:1102.1209}}].

\bibitem{Gromov:2011de}
N.~Gromov, D.~Serban, I.~Shenderovich, and D.~Volin, {\it {Quantum folded
  string and integrability: From finite size effects to Konishi dimension}},
  {\em JHEP} {\bf 1108} (2011) 046,
  [\href{http://xxx.lanl.gov/abs/1102.1040}{{\tt arXiv:1102.1040}}].

\bibitem{Beccaria:2011uz}
M.~Beccaria and G.~Macorini, {\it {Quantum folded string in $S^5$ and the
  Konishi multiplet at strong coupling}},  {\em JHEP} {\bf 1110} (2011) 040,
  [\href{http://xxx.lanl.gov/abs/1108.3480}{{\tt arXiv:1108.3480}}].

\bibitem{Gromov:2011bz}
N.~Gromov and S.~Valatka, {\it {Deeper Look into Short Strings}},
  \href{http://xxx.lanl.gov/abs/1109.6305}{{\tt arXiv:1109.6305}}.

\bibitem{Gromov:2009zb}
N.~Gromov, V.~Kazakov, and P.~Vieira, {\it {Exact Spectrum of Planar ${\cal
  N}=4$ Supersymmetric Yang-Mills Theory: Konishi Dimension at Any Coupling}},
  {\em Phys.Rev.Lett.} {\bf 104} (2010) 211601,
  [\href{http://xxx.lanl.gov/abs/0906.4240}{{\tt arXiv:0906.4240}}].

\bibitem{Frolov:2010wt}
S.~Frolov, {\it {Konishi operator at intermediate coupling}},  {\em J.Phys.A}
  {\bf A44} (2011) 065401, [\href{http://xxx.lanl.gov/abs/1006.5032}{{\tt
  arXiv:1006.5032}}].

\bibitem{Frolov:2012zv}
S.~Frolov, {\it {Scaling dimensions from the mirror TBA}},
  \href{http://xxx.lanl.gov/abs/1201.2317}{{\tt arXiv:1201.2317}}.

\bibitem{Belitsky:2003ys}
A.~V. Belitsky, A.~Gorsky, and G.~Korchemsky, {\it {Gauge / string duality for
  QCD conformal operators}},  {\em Nucl.Phys.} {\bf B667} (2003) 3--54,
  [\href{http://xxx.lanl.gov/abs/hep-th/0304028}{{\tt hep-th/0304028}}].

\bibitem{Belitsky:2006en}
A.~Belitsky, A.~Gorsky, and G.~Korchemsky, {\it {Logarithmic scaling in
  gauge/string correspondence}},  {\em Nucl.Phys.} {\bf B748} (2006) 24--59,
  [\href{http://xxx.lanl.gov/abs/hep-th/0601112}{{\tt hep-th/0601112}}].

\bibitem{Belitsky:2008mg}
A.~Belitsky, G.~Korchemsky, and R.~Pasechnik, {\it {Fine structure of anomalous
  dimensions in N=4 super Yang-Mills theory}},  {\em Nucl.Phys.} {\bf B809}
  (2009) 244--278, [\href{http://xxx.lanl.gov/abs/0806.3657}{{\tt
  arXiv:0806.3657}}].

\bibitem{Kazakov:2004nh}
V.~Kazakov and K.~Zarembo, {\it {Classical / quantum integrability in
  non-compact sector of AdS/CFT}},  {\em JHEP} {\bf 0410} (2004) 060,
  [\href{http://xxx.lanl.gov/abs/hep-th/0410105}{{\tt hep-th/0410105}}].

\bibitem{Dorey:2008zy}
N.~Dorey, {\it {A Spin Chain from String Theory}},  {\em Acta Phys.Polon.} {\bf
  B39} (2008) 3081--3116, [\href{http://xxx.lanl.gov/abs/0805.4387}{{\tt
  arXiv:0805.4387}}].

\bibitem{Dorey:2008vp}
N.~Dorey and M.~Losi, {\it {Spiky Strings and Spin Chains}},
  \href{http://xxx.lanl.gov/abs/0812.1704}{{\tt arXiv:0812.1704}}.

\bibitem{Kruczenski:2004wg}
M.~Kruczenski, {\it {Spiky strings and single trace operators in gauge
  theories}},  {\em JHEP} {\bf 0508} (2005) 014,
  [\href{http://xxx.lanl.gov/abs/hep-th/0410226}{{\tt hep-th/0410226}}].

\bibitem{Ishizeki:2008tx}
R.~Ishizeki, M.~Kruczenski, A.~Tirziu, and A.~A. Tseytlin, {\it {Spiky strings
  in $AdS_{3} \times S^{1}$ and their AdS-pp-wave limits}},  {\em Phys.Rev.}
  {\bf D79} (2009) 026006, [\href{http://xxx.lanl.gov/abs/0812.2431}{{\tt
  arXiv:0812.2431}}].

\bibitem{Jevicki:2008mm}
A.~Jevicki and K.~Jin, {\it {Solitons and AdS String Solutions}},  {\em
  Int.J.Mod.Phys.} {\bf A23} (2008) 2289--2298,
  [\href{http://xxx.lanl.gov/abs/0804.0412}{{\tt arXiv:0804.0412}}].

\bibitem{Jevicki:2009uz}
A.~Jevicki and K.~Jin, {\it {Moduli Dynamics of $AdS_{3}$ Strings}},  {\em
  JHEP} {\bf 0906} (2009) 064, [\href{http://xxx.lanl.gov/abs/0903.3389}{{\tt
  arXiv:0903.3389}}].

\bibitem{Losi:2010hr}
M.~Losi, {\it {Spiky strings and the AdS/CFT correspondence}},
  \href{http://xxx.lanl.gov/abs/1109.5401}{{\tt arXiv:1109.5401}}. Ph.D.Thesis.

\bibitem{Freyhult:2009bx}
L.~Freyhult, M.~Kruczenski, and A.~Tirziu, {\it {Spiky strings in the SL(2)
  Bethe Ansatz}},  {\em JHEP} {\bf 07} (2009) 038,
  [\href{http://xxx.lanl.gov/abs/0905.3536}{{\tt arXiv:0905.3536}}].

\bibitem{Kruczenski:2010xs}
M.~Kruczenski and A.~Tirziu, {\it {Spiky strings in Bethe Ansatz at strong
  coupling}},  {\em Phys.Rev.} {\bf D81} (2010) 106004,
  [\href{http://xxx.lanl.gov/abs/1002.4843}{{\tt arXiv:1002.4843}}].

\bibitem{Gubser:2002tv}
S.~Gubser, I.~Klebanov, and A.~M. Polyakov, {\it {A Semiclassical limit of the
  gauge / string correspondence}},  {\em Nucl.Phys.} {\bf B636} (2002) 99--114,
  [\href{http://xxx.lanl.gov/abs/hep-th/0204051}{{\tt hep-th/0204051}}].

\bibitem{Frolov:2002av}
S.~Frolov and A.~A. Tseytlin, {\it {Semiclassical quantization of rotating
  superstring in $AdS_{5}\times S^{5}$}},  {\em JHEP} {\bf 0206} (2002) 007,
  [\href{http://xxx.lanl.gov/abs/hep-th/0204226}{{\tt hep-th/0204226}}].

\bibitem{Gromov:2007aq}
N.~Gromov and P.~Vieira, {\it {The $AdS_{5}\times S^{5}$ superstring quantum
  spectrum from the algebraic curve}},  {\em Nucl.Phys.} {\bf B789} (2008)
  175--208, [\href{http://xxx.lanl.gov/abs/hep-th/0703191}{{\tt
  hep-th/0703191}}].

\bibitem{Gromov:2007ky}
N.~Gromov and P.~Vieira, {\it {Complete 1-loop test of AdS/CFT}},  {\em JHEP}
  {\bf 0804} (2008) 046, [\href{http://xxx.lanl.gov/abs/0709.3487}{{\tt
  arXiv:0709.3487}}].

\bibitem{Gromov:2008ec}
N.~Gromov, S.~Schafer-Nameki, and P.~Vieira, {\it {Efficient precision
  quantization in AdS/CFT}},  {\em JHEP} {\bf 0812} (2008) 013,
  [\href{http://xxx.lanl.gov/abs/0807.4752}{{\tt arXiv:0807.4752}}].

\bibitem{Basso:2011rs}
B.~Basso, {\it {An exact slope for AdS/CFT}},
  \href{http://xxx.lanl.gov/abs/1109.3154}{{\tt arXiv:1109.3154}}.

\bibitem{Arutyunov:2009ax}
G.~Arutyunov, S.~Frolov, and R.~Suzuki, {\it {Exploring the mirror TBA}},  {\em
  JHEP} {\bf 1005} (2010) 031, [\href{http://xxx.lanl.gov/abs/0911.2224}{{\tt
  arXiv:0911.2224}}].

\bibitem{Kruczenski:2008bs}
M.~Kruczenski and A.~A. Tseytlin, {\it {Spiky strings, light-like Wilson loops
  and pp-wave anomaly}},  {\em Phys.Rev.} {\bf D77} (2008) 126005,
  [\href{http://xxx.lanl.gov/abs/0802.2039}{{\tt arXiv:0802.2039}}].

\bibitem{Beisert:2003ea}
N.~Beisert, S.~Frolov, M.~Staudacher, and A.~A. Tseytlin, {\it {Precision
  spectroscopy of AdS / CFT}},  {\em JHEP} {\bf 0310} (2003) 037,
  [\href{http://xxx.lanl.gov/abs/hep-th/0308117}{{\tt hep-th/0308117}}].

\bibitem{SchaferNameki:2010jy}
S.~Schafer-Nameki, {\it {Review of AdS/CFT Integrability, Chapter II.4: The
  Spectral Curve}},  \href{http://xxx.lanl.gov/abs/1012.3989}{{\tt
  arXiv:1012.3989}}.

\bibitem{Methods}
C.~M. Bender and S.~A. Orszag, {\em Advanced Mathematical Methods for
  Scientists and Engineers}.
\newblock Mc Graw-Hill, 1984.

\bibitem{Beccaria:2012tu}
M.~Beccaria and G.~Macorini, {\it {Resummation of semiclassical short folded
  string}},  \href{http://xxx.lanl.gov/abs/1201.0608}{{\tt arXiv:1201.0608}}.

\end{thebibliography}\endgroup
\bibliographystyle{JHEP}

\end{document}